\documentclass[conference]{IEEEtran}
\ifCLASSINFOpdf
\else
   \usepackage[dvips]{graphicx}
  \DeclareGraphicsExtensions{.eps}
\fi
\usepackage[cmex10]{amsmath}
\usepackage{amsfonts,amssymb}
\usepackage[tight,footnotesize]{subfigure}
\newcommand{\Cr}{{\bf C}}     \newcommand{\Ct}{\tilde{{\bf C}}}
\newcommand{\Crl}{\Cr^{(l)}}  \newcommand{\Ctl}{\Ct^{(l)}}
\newcommand{\Crj}{\Cr^{(j)}}  \newcommand{\Ctj}{\Ct^{(j)}}
\newcommand{\Crk}{\Cr^{(k)}}  \newcommand{\Ctk}{\Ct^{(k)}}
\newcommand{\dr}{\delta}      \newcommand{\dt}{\tilde{\delta}}
\newcommand{\drl}{\delta_l}   \newcommand{\dtl}{\tilde{\delta}_l}
\newcommand{\frl}{f_l}        \newcommand{\ftl}{\tilde{f}_l}
\newcommand{\bs}{\boldsymbol}

\newtheorem{proposition}{Proposition}
\newtheorem{lemma}{Lemma}
\renewenvironment{IEEEproof}
{\noindent\textbf{Proof:}}{\hfill$\square$\smallbreak} 

\newcounter{MYtempeqncnt} %

\begin{document}
\title{On the Capacity Achieving Covariance Matrix for
    Frequency Selective  MIMO Channels\\ Using the Asymptotic Approach}

\author{\IEEEauthorblockN{Florian Dupuy}
\IEEEauthorblockA{Thales Communication EDS/SPM\\
160 Bd de Valmy\\
92704 Colombes Cedex, France}
\and
\IEEEauthorblockN{Philippe Loubaton}
\IEEEauthorblockA{Universit\'e de Paris-Est\\
IGM LabInfo, UMR-CNRS 8049\\
5, Bd Descartes, Champs-sur-Marne\\
77454 Marne-la-Vall\'ee Cedex 2, France}}

\maketitle

\begin{abstract}
  In this contribution, an algorithm for evaluating 
  the capacity-achieving input covariance
  matrices for frequency selective Rayleigh MIMO channels is proposed. In contrast 
  with the flat fading Rayleigh cases, no closed-form expressions for the eigenvectors of the
  optimum input covariance matrix are available. Classically, both the
  eigenvectors and eigenvalues are computed numerically and the corresponding 
  optimization algorithms remain computationally very demanding.
  
  In this paper, it is proposed to optimize (w.r.t. the input covariance matrix) a large system approximation of the average mutual information 
  derived by Moustakas and Simon. 
  An algorithm based on an iterative water filling scheme is proposed, and its convergence is studied.  Numerical simulation results show that, even
  for a moderate number of transmit and receive antennas, the new
  approach provides the same results as direct maximization approaches
  of the average mutual information. \end{abstract}
\IEEEpeerreviewmaketitle

\section{Introduction}
When the channel state information is available at both the 
receiver and the transmitter of a MIMO system, the problem
of designing the transmitter in order to maximize the (Gaussian) mutual information 
of the system has been addressed successfully in a number of papers. 
This problem is however more difficult when the transmitter has the knowledge 
of the statistical properties of the channel, a more realistic assumption 
in the context of mobile systems. In this case, the mutual information 
is replaced by the average mutual information (EMI), which, of course, 
is more complicated to optimize. \medskip

The optimization problem of the EMI has been addressed extensively in the case
of certain flat fading Rayleigh channels. In the context of the so-called
Kronecker model, it has been shown by various authors (see e.g.
\cite{Goldsmith-Jafar-etal-03} for a review) that the eigenvectors of
the optimal input covariance matrix must coincide with the eigenvectors of
the transmit correlation matrix. It is therefore sufficient to
evaluate the eigenvalues of the optimal matrix, a problem which can be
solved by using standard optimization algorithms. Similar results 
have been obtained for flat fading uncorrelated Rician channels (\cite{Hoesli-Kim-Lapidoth-05}).\medskip

In this paper, we consider this EMI maximization problem in the case of 
popular frequency selective MIMO channels (see e.g. \cite{boelcskei}, \cite{Moustakas-Simon-07}) with independent 
paths. In this context, the eigenvectors of the optimum transmit covariance matrix 
have no closed expressions, so that both the eigenvalues and the eigenvectors of the 
matrix have to be evaluated numerically. For this, it is possible to adapt the 
approach of \cite{Vu-Paulraj-05} developed in the context of correlated 
Rician channels. However, the corresponding algorithms are computationally very demanding as
they heavily rely on intensive Monte-Carlo simulations. We therefore propose to optimize the approximation
of the EMI, derived by Moustakas and Simon (\cite{Moustakas-Simon-07}), in principle valid when the number of transmit and receive antennas 
converge to infinity at the same rate, but accurate for realistic numbers of antennas.  This will turn out to be a simpler problem. 
We mention that, while \cite{Moustakas-Simon-07} contains some results related to the structure of the 
argument of the maximum of the EMI approximation, \cite{Moustakas-Simon-07} does not propose any optimization algorithm.\medskip

We first review the results of \cite{Moustakas-Simon-07} related to the large system approximation of the EMI. 
The expression of the approximation depends on the solutions of a non linear system. The existence
and the uniqueness of the solutions is not addressed in \cite{Moustakas-Simon-07}. As our optimization algorithm
needs to solve this system, we clarify this crucial point.
Next, we present our maximization algorithm of 
the EMI approximation. It is based on an iterative waterfilling  algorithm which, in some sense, can be seen as a
generalization of \cite{Wen-Com-06} devoted to the
Rayleigh context and of \cite{dum-hac-las-lou-naj-09} devoted to the correlated Rician case: each iteration will be devoted to solve 
the above mentioned  system of nonlinear equations as well as a standard waterfilling problem. It is proved 
that the algorithm converges towards the optimum
input covariance matrix as long as it converges\footnote{Note however
  that we have been unable to prove formally its convergence.}.\medskip

The paper is organized as follows. Section \ref{sec:model} is devoted
to the presentation of the channel model, the underlying
assumptions, the problem statement.
The maximization problem of the EMI approximation is studied in section \ref{sec:algo}.
Numerical results are provided in section \ref{sec:simulations}.

\section{Problem statement}
\label{sec:model}

\subsection{General Notations} In this paper, the notations $s$, ${\bf
x}$, ${\bf M}$, stand for scalars, vectors and matrices, respectively.
As usual, $\|{\bf x} \|$ represents the Euclidian norm of vector ${\bf
  x}$, and $\| {\bf M} \|$, $\rho({\bf M})$ and $| {\bf M} |$ respectively stand for the spectral norm, 
the spectral radius and the determinant of matrix ${\bf
  M}$. The superscripts $(.)^T$ and $(.)^H$ represent respectively the
transpose and transpose conjugate. The trace of ${\bf M}$ is denoted by
$\mathrm{Tr}({\bf M})$. The mathematical expectation operator is
denoted by $\mathbb{E}(\cdot)$ 

All along this paper, $r$ and $t$ stand for the number of receive and
transmit antennas.  Certain quantities will be studied in the
asymptotic regime $t \to \infty$, $r \to \infty$ in
such a way that%
$t/r \to c \in (0,\infty)$. In order to
simplify the notations, $t \to \infty$ should be understood
from now on as $t \to \infty$, $r \to \infty$ and
$t/r \to c \in (0,\infty)$.

Several variables used throughout this paper depend on various parameters, e.g. 
the number of antennas, the noise level, the covariance matrix of the transmitter, etc.
In order to simplify the notations, we may not always mention all these dependencies.

\begin{figure*}[!t]
\normalsize
\setcounter{MYtempeqncnt}{\value{equation}}
\setcounter{equation}{6}
\begin{equation}
  \label{eq:expre-Ibarre}
  \overline{I}({\bf Q}) = \log \left| {\bf I} + \sum_{l=1}^{L} \dtl({\bf Q}) \Crl \right|
   +\log \left|{\bf I} + {\bf Q} \left( \sum_{l=1}^{L} \drl({\bf Q}) \Ctl \right) \right|
   -\sigma^{2} t \left( \sum_{l=1}^{L}  \drl({\bf Q})  \dtl({\bf Q}) \right)
\end{equation} 
\setcounter{equation}{\value{MYtempeqncnt}}
\hrulefill
\end{figure*}

\subsection{Channel model}

We consider a wireless MIMO link with $t$ transmit and $r$ receive
antennas corrupted by a multi-paths propagation channel. The discrete-time propagation channel between the transmitter and the receiver 
is characterized by the input-output equation
\begin{equation}
{\bf y}(n) = \sum_{l=1}^{L} {\bf H}_l {\bf s}(n-l+1) + {\bf n}(n)= [{\bf H}(z)]{\bf s}(n) + {\bf n}(n)
\label{eq:model-signaux}
\end{equation}
where ${\bf s}(n) = (s_1(n), \ldots, s_t(n))^{T}$  
represents the transmit vector at time $n$, ${\bf y}(n) = (y_1(n), \ldots, y_r(n))^{T}$ the receive vector,
and where ${\bf n}(n)$ is an additive Gaussian noise such that $\mathbb{E}({\bf n}(n){\bf n}(n)^{H}) = \sigma^{2} {\bf I}$. ${\bf H}(z)$ denotes the transfer function of the discrete-time equivalent channel defined by
\begin{equation}
\label{eq:def-transfert}
{\bf H}(z) = \sum_{l=1}^{L} {\bf H}_l \, z^{-(l-1)}
\end{equation}
Each coefficient ${\bf H}_l$ is assumed to be a Gaussian random matrix given by
\begin{equation}
\label{eq:structure-Hl}
{\bf H}_l = \frac{1}{\sqrt{t}} (\Crl)^{1/2} {\bf W}_l (\Ctl)^{1/2}
\end{equation}
where ${\bf W}_l$ is a $r \times t$ random matrix whose entries are independent 
and identically distributed complex circular Gaussian random variables, with zero mean and unit variance. The matrices 
$\Crl$ and $\Ctl$ are positive definite, and account for the receive and transmit 
antenna correlation. We also assume that for each $k \neq l$, matrices ${\bf H}_k$ and
${\bf H}_l$ are independent.

In the context of this paper, the channel matrices are assumed perfectly known at the
receiver side. However, only the statistics of the $({\bf H}_l)_{l=1, \ldots, L}$, i.e. 
matrices $(\Ctl, \Crl)_{l=1, \ldots, L}$, are available at the transmitter side. 

\subsection{Ergodic capacity of the channel.}
Let ${\bf Q}(e^{2 i \pi \nu})$ be the $t \times t$ spectral density matrix 
of the transmit signal ${\bf s}(n)$, which is assumed to verify the transmit power 
condition
\begin{equation}
\label{eq:power}
\frac{1}{t} \int_0^1 \mathrm{Tr}({\bf Q}(e^{2 i \pi \nu})) d \nu = 1
\end{equation}
Then, the (Gaussian) ergodic mutual information $I({\bf Q}(.))$ between the transmitter and the receiver is defined as
\begin{equation}
\label{eq:def-EMI}
I({\bf Q}(.)) = \mathbb{E}_{\cal{W}} \left[ \int_0^1 \log \left| {\bf I}_r + \frac{1}{\sigma^{2}} {\bf H}(.) {\bf Q}(.) {\bf H}(.)^{H} \right| \, d\nu \right]
\end{equation}
where $\mathbb{E}_{\cal{W}}[.]=\mathbb{E}_{({\bf W}_l)_{l=1, \ldots, L}}[.]$. The ergodic capacity of the MIMO channel is equal to the maximum of $I({\bf Q}(.))$ over the set of all spectral density matrices 
satisfying the constraint (\ref{eq:power}). The hypotheses formulated on the statistics of the channel allow however to 
limit the optimization to the set of positive matrices which are independent of the frequency $\nu$. This is because the 
probability distribution of matrix ${\bf H}(e^{2 i \pi \nu})$ is clearly independent of the frequency $\nu$. More precisely, the mutual 
information ${\bf I}({\bf Q}(.))$ is also given by 
\[
I({\bf Q}(.)) = \mathbb{E}_{{\bf H}} \left[ \int_0^1 \log \left| {\bf I}_r + \frac{1}{\sigma^{2}} {\bf H}(1) {\bf Q}(.) {\bf H}(1)^{H} \right| \, d\nu \right]
\]
where ${\bf H} = \sum_{l=1}^{L} {\bf H}_l={\bf H}(1)$. Using the concavity of the logarithm, we obtain that
\[
I({\bf Q}(.))  \leq  \mathbb{E}_{{\bf H}} \left[ \log \left| {\bf I}_r + \frac{1}{\sigma^{2}} {\bf H}(1) \left( \int_0^1 {\bf Q}(.) d\nu \right) \, {\bf H}(1)^{H} \right| \,  \right]
\]
We denote by ${\cal C}$ the cone of non negative hermitian matrices, and by ${\cal C}_1$ the subset of all matrices ${\bf Q}$ of ${\cal C}$ satisfying 
$\frac{1}{t} \mathrm{Tr}({\bf Q}) = 1$. If ${\bf Q}$ is an element of ${\cal C}_1$, the mutual information $I({\bf Q})$ reduces to  
\begin{equation}
I({\bf Q}) =  \mathbb{E}_{{\bf H}} \left[\log \left| {\bf I}_r + \frac{1}{\sigma^{2}} {\bf H} {\bf Q}  {\bf H}^{H} \right| \right]
\label{eq:expre-finale-I} 
\end{equation}
It is strictly concave on the convex set ${\cal C}_1$ and reaches its maximum at a unique element ${\bf Q}_* \in {\cal C}_1$. It is clear that 
if ${\bf Q}(e^{2 i \pi \nu})$ is any spectral density satisfying (\ref{eq:power}), then the matrix $ \int_0^1 {\bf Q}(e^{2 i \pi \nu}) d\nu $ 
is an element of ${\cal C}_1$. Therefore, 
\[
I({\bf Q}(.)) \leq I({\bf Q}_{*})
\]
for each spectral density matrix verifying  (\ref{eq:power}). This shows that the maximum of function $I$ over the set of all spectral densities satisfying (\ref{eq:power}) is reached on the set ${\cal C}_1$. The 
ergodic capacity ${\cal C}_E$ of the channel is thus equal to 
\[
{\cal C}_E  = \max_{{\bf Q} \in {\cal C}_1} I({\bf Q})
\]
If the matrices $(\Crl)_{l=1, \ldots, L}$ coincide with a matrix ${\bf C}$, matrix ${\bf H}$ follows a Kronecker model with transmit and
receive covariance matrices $\sum_{l=1}^{L} \Ctl$ and ${\bf C}$ respectively \cite{Cedric-GC}. In this case, 
the eigenvectors of the optimum matrix ${\bf Q}_*$ coincide with the eigenvectors of $\sum_{l=1}^{L} \Ctl$. 
The situation is similar if the transmit covariance matrices $(\Ctl)_{l=1, \ldots, L}$ coincide. In the most general case, 
the eigenvectors of ${\bf Q}_*$ have however no closed form expression. The evaluation of ${\bf Q}_*$ and of the channel 
capacity ${\cal C}_E$ is thus a more difficult problem. A possible solution consists in adapating the 
Vu-Paulraj approach (\cite{Vu-Paulraj-05}) to the present context. However, the algorithm presented in \cite{Vu-Paulraj-05}
is very demanding since the evaluation of the gradient and the Hessian of $I({\bf Q})$ requires intensive Monte-Carlo simulations. 

\subsection{The large system approximation of $I({\bf Q})$.}
\addtocounter{equation}{1}%
When $t$ and $r$ converge to $\infty$ while
$t/r \rightarrow c$, $c \in (0, \infty)$, 
\cite{Moustakas-Simon-07} showed that $I({\bf Q})$ can be approximated by $\overline{I}({\bf Q})$ defined by
(\ref{eq:expre-Ibarre}) at the top of the page,
where $(\dr_1({\bf Q}),\ldots, \dr_L({\bf Q}))^T={\bs \dr({\bf Q})}$ and $(\dt_1({\bf Q}), \ldots,\dt_L({\bf Q}))^T={\bs \dt({\bf Q})}$ are the positive solutions of the system of $2L$ equations:
\begin{equation}
\left\{\begin{array}{l}
\kappa_l = \frl(\tilde{\bs{\kappa}}) \vspace{1mm}\\
\tilde{\kappa}_l = \ftl(\bs{\kappa}, {\bf Q})
\end{array}\right.
\label{eq:canonique}
\end{equation}
with $\bs{\kappa} = (\kappa_1, \ldots, \kappa_L)^{T}$ and  $\tilde{\bs{\kappa}} = (\tilde{\kappa}_1, \ldots, \tilde{\kappa}_L)^{T}$, and 
\begin{equation}
\label{eq:canoniquebis}
\left\{\begin{array}{l}
  \frl(\tilde{\bs{\kappa}}) = \frac{1}{t} \mathrm{Tr} \left[ \Crl {\bf T}(\tilde{\bs{\kappa}}) \right]
  \vspace{1mm}\\
  \ftl(\bs{\kappa}, {\bf Q}) = \frac{1}{t} \mathrm{Tr} \left[ {\bf Q}^{1/2} \Ctl {\bf Q}^{1/2}  \tilde{\bf T}(\bs{\kappa}, {\bf Q}) \right]
\end{array} \right.
\end{equation}
where
\begin{equation}
\label{eq:canoniqueter}
\left\{\begin{array}{l}
{\bf T}^{-1}(\tilde{\bs{\kappa}})=\sigma^{2} \left({\bf I} + \sum_{j=1}^{L} \tilde{\kappa}_j \Crj \right)
\vspace{1mm}\\ \tilde{\bf T}^{-1}(\bs{\kappa}, {\bf Q})=\sigma^{2} \left( {\bf I} + \sum_{j=1}^{L} \kappa_j  {\bf Q}^{1/2} \Ctj  {\bf Q}^{1/2} \right)
\end{array} \right.
\end{equation}
\cite{Moustakas-Simon-07} is based on the replica method, a useful and simple trick whose mathematical relevance is not yet proved in the present
context. However, using large random matrix technics similar to those of (\cite{hac-kor-lou-naj-pas-08}, \cite{dum-hac-las-lou-naj-09}), it is possible to prove rigorously that, under mild technical extra assumptions,
 $\overline{I}({\bf Q}) = I({\bf Q}) + O(\frac{1}{t})$. This point is outside the scope of the present paper. 

We also mention that \cite{Moustakas-Simon-07} assumed implicitely the existence and the uniqueness of positive solutions of (\ref{eq:canonique}) without justification. We therefore precise this important point, and
also show that  $(\drl({\bf Q}))_{l=1, \ldots, L}$ and $(\dtl({\bf Q}))_{l=1, \ldots, L}$ can be evaluated using a fixed point algorithm. \medskip

\subsubsection{Existence}
Using analytic continuation technique and results of \cite{HLN07},  it can be shown that the following fixed point algorithm, 
initialized as follows, converges:
	\begin{itemize}
	\item Initialization: $\delta_l^{(0)}>0$,  $\tilde{\delta}_l^{(0)}>0$, $l=1,\ldots,L$.
	\item Evaluation of the $\drl^{(n+1)}$ and $\dtl^{(n+1)}$ from ${\bs{\delta}}^{(n)}=(\delta_1^{(n)}, \ldots, \delta_L^{(n)})^T$ and $\tilde{\bs{\delta}}^{(n)}=(\tilde{\delta}_1^{(n)}, \ldots, \tilde{\delta}_L^{(n)})^{T}$:
	\begin{eqnarray}
	\label{eq:point-fixe}
	\left\{\begin{array}{l}
	\delta_l^{(n+1)} = \frl(\tilde{\bs{\delta}}^{(n)}) \vspace{1mm}\\
	\tilde{\delta}_l^{(n+1)} = \ftl(\bs{\delta}^{(n)}, {\bf Q})
	\end{array}\right.
	\end{eqnarray}
	\end{itemize}
	Besides, it can be proved that the limit of $({\bs{\delta}}^{(n)},\tilde{\bs{\delta}}^{(n)})$ when $n\rightarrow\infty$
	satisfies equation (\ref{eq:canonique}), and that all the entries of this limit are positive.
	Hence, the convergence of the algorithm yields the existence of a solution to (\ref{eq:canonique}).
\subsubsection{Uniqueness}
	In order to simplify the notations, we consider in this part the case ${\bf Q} = {\bf I}$.
	In order to address the general case, it is sufficient to change matrices $(\Ctl)_{l=1, \ldots, L}$ into  $({\bf Q}^{1/2}\Ctl {\bf Q}^{1/2})_{l=1, \ldots, L}$ in what follows.
	Let $(\bs{\dr},\bs{\dt})$ and $(\bs{\dr}',\bs{\dt}')$ be two solutions of the canonical equation (\ref{eq:canonique}). We denote $({\bf T}, \tilde{\bf T})$ and $({\bf T}', \tilde{\bf T}')$ the associated matrices defined by (\ref{eq:canoniqueter}). Introducing ${\bf e}=\bs{\dr}-\bs{\dr}'=(e_1,\ldots,e_L)^T$ we have:
	\begin{align}
	e_l &= \frac{1}{t}\mathrm{Tr}\left[\Crl{\bf T}({\bf T'}^{-1}-{\bf T}^{-1}){\bf T'}\right] \notag
	\\ &= \frac{\sigma^2}{t}\sum_{k=1}^L(\tilde{\delta}_k'-\tilde{\delta}_k)\mathrm{Tr}\left(\Crl{\bf T}\Crk{\bf T'}\right) \label{el:eqn}
	\end{align}
	Similarly, with $\tilde{\bs{e}}=\bs{\dt}-\bs{\dt}'=(\tilde{e}_1,\ldots,\tilde{e}_L)^T$,
	\begin{equation}
	\tilde{e}_k = \frac{\sigma^2}{t}\sum_{l=1}^L(\delta_l'-\delta_l)\mathrm{Tr}\left(\Ctk\tilde{\bf T}\Ctl\tilde{\bf T'}\right) \label{elt:eqn}
	\end{equation}
	And (\ref{el:eqn}) and (\ref{elt:eqn}) can be written together as
	\begin{equation}
	\left[\begin{array}{cc} {\bf I} & \sigma^2{\bf A}({\bf T},{\bf T'}) \\ \sigma^2\tilde{\bf A}(\tilde{\bf T},\tilde{\bf T'}) & {\bf I}  \end{array}\right] \left[\begin{array}{c}{\bf e} \\ \tilde{\bf e} \end{array}\right]= {\bf 0} \label{unq:matrix}
	\end{equation}
	with ${\bf A}_{kl}({\bf T},{\bf T'})=\frac{1}{t}\mathrm{Tr}\left(\Crk{\bf T}\Crl{\bf T'}\right)$ and $\tilde{\bf A}_{kl}(\tilde{\bf T},\tilde{\bf T'})=\frac{1}{t}\mathrm{Tr}(\Ctk\tilde{\bf T}\Ctl\tilde{\bf T'})$.
	We will now prove that $\rho({\bf M})<1$, with ${\bf M}=\sigma^4\tilde{\bf A}(\tilde{\bf T},\tilde{\bf T}'){\bf A}({\bf T},{\bf T'})$. This will imply that matrix ${\bf M}$ is invertible, and thus that ${\bf e}=\tilde{\bf e}={\bf 0}.$
	\begin{align}
	|{\bf M}&_{kl}| = \left|\frac{\sigma^4}{t^2}\sum_{j=1}^L\mathrm{Tr}(\Ctk\tilde{\bf T}\Ctj\tilde{\bf T'})\mathrm{Tr}(\Crj{\bf T}\Crl{\bf T'})\right| \notag
	\\ &\leq \frac{\sigma^4}{t^2}\sum_{j=1}^L\left|\mathrm{Tr}(\Ctk\tilde{\bf T}\Ctj\tilde{\bf T'})\right|\left|\mathrm{Tr}(\Crj{\bf T}\Crl{\bf T'})\right|
	\label{Mkl-ineq}
	\end{align}
	Thanks to the inequality $|\mathrm{Tr}({\bf AB})|\leq\sqrt{\mathrm{Tr}({\bf AA^H})\mathrm{Tr}({\bf BB^H})}$, we have
	\begin{align}
	&\frac{1}{t}\left|\mathrm{Tr}(\Ctk\tilde{\bf T}\Ctj\tilde{\bf T'})\right| \leq 
	\sqrt{\tilde{\bf A}_{kj}(\tilde{\bf T},\tilde{\bf T})\tilde{\bf A}_{kj}(\tilde{\bf T}',\tilde{\bf T'})}
	\label{1stTr}
	\\
	&\frac{1}{t}\left|\mathrm{Tr}(\Crj{\bf T}\Crl{\bf T'})\right|
	\leq \sqrt{{\bf A}_{jl}({\bf T},{\bf T}){\bf A}_{jl}({\bf T}',{\bf T'})}
	\label{2ndTr}
	\end{align}
	Using (\ref{1stTr}) and (\ref{2ndTr}) in (\ref{Mkl-ineq}) gives
	\[
	|{\bf M}_{kl}| \leq \sigma^4\sum_{j=1}^L\sqrt{\tilde{\bf A}_{kj}(\tilde{\bf T})\tilde{\bf A}_{kj}(\tilde{\bf T'}){\bf A}_{jl}({\bf T}){\bf A}_{jl}({\bf T}')}
	\]
	with $\tilde{\bf A}(\tilde{\bf T})=\tilde{\bf A}(\tilde{\bf T},\tilde{\bf T})$ and ${\bf A}({\bf T})={\bf A}({\bf T},{\bf T})$. And, using Cauchy-Schwarz inequality,
	\begin{align}
	|{\bf M}_{kl}| &\leq \sigma^4\sqrt{\bigg(\sum_{j=1}^L\tilde{\bf A}_{kj}(\tilde{\bf T}){\bf A}_{jl}({\bf T})\bigg)\bigg(\sum_{j=1}^L\tilde{\bf A}_{kj}(\tilde{\bf T'}){\bf A}_{jl}({\bf T'})\bigg)} \notag
	\end{align}
	Hence, we have $|{\bf M}_{kl}|\leq{\bf P}_{kl}$ $\forall k,l$, where the matrix $\bf{P}$ is defined by
	${\bf P}_{kl}=\sqrt{(\sigma^4\tilde{\bf A}(\tilde{\bf T}){\bf A}({\bf T}))_{kl}}\sqrt{(\sigma^4\tilde{\bf A}(\tilde{\bf T'}){\bf A}({\bf T}'))_{kl}}$.
	Theorem 8.1.18 of \cite{horn1990matrix} then yields $\rho({\bf M})\leq\rho({\bf P})$.
	Besides, Lemma 5.7.9 of \cite{horn1994topics} used on the definition of ${\bf P}$ gives:
	\[
	\rho({\bf P})\leq\sqrt{\rho\left(\sigma^4\tilde{\bf A}(\tilde{\bf T}){\bf A}({\bf T})\right)}\sqrt{\rho\left(\sigma^4\tilde{\bf A}(\tilde{\bf T'}){\bf A}({\bf T}')\right)}
	\]
	We now introduce the following lemma: \medskip
	\begin{lemma}
	$\rho\left(\sigma^4\tilde{\bf A}(\tilde{\bf T}){\bf A}({\bf T})\right)<1$
	\label{lemAAt}
	\end{lemma} \medskip
	\begin{IEEEproof}
	The $\delta_l$ can be written as:
	\begin{align}
	\delta_l &=\frac{1}{t}\mathrm{Tr}(\Crl{\bf T}{\bf T}^{-1}{\bf T}) \notag
	\\ &=\frac{\sigma^2}{t}\mathrm{Tr}(\Crl{\bf T}{\bf T})+\frac{\sigma^2}{t}\sum_{k=1}^L\tilde{\delta_k}\mathrm{Tr}(\Crl{\bf T}\Crk{\bf T})  \notag
	\end{align}
	And
	$\tilde{\delta}_l=\frac{\sigma^2}{t}\mathrm{Tr}(\Ctl\tilde{\bf T}\tilde{\bf T})+\frac{\sigma^2}{t}\sum_{k=1}^L{\delta_k}\mathrm{Tr}(\Ctl\tilde{\bf T}\Ctk\tilde{\bf T})$ similarly, thus:
	\[
	\left[\begin{array}{c} {\bs \delta} \\ \tilde{\bs \delta} \end{array}\right]
	= \sigma^2\left[\begin{array}{cc} 0 & {\bf A}({\bf T}) \\ \tilde{\bf A}(\tilde{\bf T}) & 0 \end{array}\right]
	\left[\begin{array}{c} {\bs \delta} \\ \tilde{\bs \delta} \end{array}\right]
	+ {\bf v}
	\]
	This equality is of the form ${\bf u}={\bf B}{\bf u}+{\bf v}$, where the entries of ${\bf u}$ and ${\bf v}$ are positive, and where the entries of ${\bf B}$ are non-negative. A direct application of Corollary 8.1.29 of \cite{horn1990matrix} then implies $\rho({\bf B}) \leq 1-\frac{\min v_l}{\max u_l}< 1$. Noticing that $(\rho({\bf B}))^2=\rho(\sigma^4\tilde{\bf A}(\tilde{\bf T}){\bf A}({\bf T}))$ ends the proof of Lemma \ref{lemAAt}.
	\end{IEEEproof}
	
	We also have of course $\rho(\sigma^4\tilde{\bf A}(\tilde{\bf T'}){\bf A}({\bf T'}))<1$, so that finally: \[\rho({\bf M})\leq\rho({\bf P})<1.\]

\section{Maximization algorithm}
\label{sec:algo}
Using the same methods as \cite{dumont-2007} section IV, the approximation $\overline{I}({\bf Q})$ can be shown to be a strictly concave function over the compact set ${\cal C}_1$. Therefore it admits a unique argmax that we denote $\overline{{\bf Q}}_*$. As 
${\cal C}_1$ is convex, it is well known that $\overline{{\bf Q}}_*$ is characterized by the property 
\begin{equation}
\label{eq:carac-max}
\langle\nabla \overline{I}(\overline{{\bf Q}}_*), {\bf P} - \overline{{\bf Q}}_*\rangle \leq 0
\end{equation}
for each matrix ${\bf P} \in {\cal C}_1$, where $\langle\nabla \overline{I}({\bf Q}), {\bf P} - {\bf Q}\rangle$ represents 
the limit of $\lambda^{-1} \left(\overline{I}({\bf Q}+ \lambda({\bf P} -{\bf Q})) - \overline{I}({\bf Q})\right)$ when $\lambda \rightarrow 0$, $\lambda > 0$. 
We now consider the function ${\cal V}({\bf Q},{\bs \kappa},\tilde{\bs \kappa})$ defined by:
\begin{equation}\begin{split}
   {\cal V}({\bf Q},{\bs \kappa},\tilde{\bs \kappa}) = &\log | {\bf I} + {\bf C}(\tilde{\bs \kappa}) |
   +\log |{\bf I} + {\bf Q} \tilde{\bf C}({\bs \kappa}) |
   \\&-\sigma^{2} t \sum_{l=1}^{L}  \kappa_l \tilde\kappa_l
\end{split}\label{def:V}\end{equation}
where
$\tilde{\bf C}({\bs \kappa})=\sum_{l=1}^{L} \kappa_l \Ctl$ and ${\bf C}(\tilde{\bs \kappa})=\sum_{l=1}^{L} \tilde\kappa_l \Crl$. 
Note that we have ${\cal V}({\bf Q},{\bs \delta}({\bf Q}),\tilde{\bs \delta}({\bf Q}))=\overline{I}({\bf Q})$. We have then the following result:\medskip
\begin{proposition}
\label{prop:waterfilling}
 Denote $\bs{\delta}_*={\bs \delta}({\bf Q}_*)$ and $\tilde{\bs{\delta}}_*=\tilde{\bs \delta}({\bf Q}_*)$. Matrix ${\bf Q_*}$ is the solution of the standard waterfilling problem: maximize over ${\bf Q \in {\cal C}_1}$ the function $\log|{\bf I}+{\bf Q}\tilde{\bf C}(\bs{\delta}_*)|$.
\end{proposition} \medskip
\begin{IEEEproof}
Due to lack of space, only the key points are given.
We first remark that maximizing function ${\bf Q}\mapsto\log|{\bf I}+{\bf Q}\tilde{\bf C}(\bs{\delta}_*)|$ is equivalent to maximizing function ${\bf Q}\mapsto{\cal V}({\bf Q},\bs{\delta}_*,\tilde{\bs{\delta}}_*)$ by (\ref{def:V}).
The proof is then based on the observation that
\begin{align}
\label{eq:derivee-kappa}
 &\frac{\partial {\cal V}}{\partial \kappa_l}=-\sigma^2t\big(\ftl(\bs{\kappa}, {\bf Q})-\tilde{\kappa}_l\big)
 \\
& \frac{\partial {\cal V}}{\partial \tilde{\kappa}_l}=-\sigma^2t\big(\frl(\tilde{\bs{\kappa}})-\kappa_l\big)
\label{eq:derivee-tildekappa}
\end{align}
are zero at point $(\bs{\delta}({\bf Q}), \tilde{\bs{\delta}}({\bf Q}))$. This implies that for each ${\bf P} \in {\cal C}_1$, $\langle\nabla \overline{I}(\overline{{\bf Q}}_*), {\bf P} - \overline{{\bf Q}}_*\rangle$ coincides 
with $\langle\nabla_{{\bf Q}} {\cal V}(\overline{{\bf Q}}_*,\bs{\delta}_*, \tilde{{\bs \delta}}_*), {\bf P} - \overline{{\bf Q}}_*\rangle$.
As function ${\bf Q} \rightarrow {\cal V}({\bf Q}, \bs{\delta}_*, \tilde{{\bs \delta}}_*)$ is strictly concave on 
${\cal C}_1$, (\ref{eq:carac-max}) implies that its argmax on ${\cal C}_1$ coincides with $\overline{{\bf Q}}_*$.
\end{IEEEproof}

Proposition \ref{prop:waterfilling} shows that the optimum matrix is solution of a waterfilling problem associated to 
the covariance matrix $\tilde{\bf C}(\bs{\delta}_*)$. Although this result provides some insight on the structure of 
$\overline{{\bf Q}}_{*}$, it cannot be used to evaluate it because matrix  $\tilde{\bf C}(\bs{\delta}_*)$ depends itself of $\overline{{\bf Q}}_{*}$. 
We now introduce an optimization algorithm of $\overline{I}({\bf Q})$; the iterative scheme is the following:
\begin{itemize}
	\item Initialization: ${\bf Q}_0={\bf I}$%
	\item Evaluation of ${\bf Q}_k$ from%
	${\bf Q}_{k-1}$: $({\bs{\delta}}^{(k)}, \tilde{\bs{\delta}}^{(k)})$ is defined as the unique solution of (\ref{eq:canonique}) in which ${\bf Q}={\bf Q}_{k-1}$. Then ${\bf Q}_k$ is defined as the maximum of function ${\bf Q}\mapsto\log\left|{\bf I}+{\bf Q}\tilde{\bf C}(\bs{\delta}^{(k)})\right|$ on ${\cal C}_1$.
\end{itemize}

\medskip
We now establish a result which shows that if the algorithm converges, then it converges towards $\overline{{\bf Q}}_*$. 
\medskip
\begin{proposition}
\label{prop:convergence}
Assume that 
\begin{equation}
\label{eq:hypothese}
\lim_{k \rightarrow \infty} {\bs{\delta}}^{(k)} - {\bs{\delta}}^{(k-1)} =\lim_{k \rightarrow \infty} {\tilde{\bs{\delta}}}^{(k)} - \tilde{{\bs{\delta}}}^{(k-1)} =  0
\end{equation} 
Then, the algorithm converges torwards matrix $\overline{{\bf Q}}_*$. 
\end{proposition}
\medskip
\begin{IEEEproof}
Due to the lack of space, we just outline the proof which is similar to the proof of Proposition 
6 of \cite{dum-hac-las-lou-naj-09}. As ${\cal C}_1$ is compact, we have just to verify that 
each convergent subsequence $({\bf Q}_{\psi(k)})_{k\in\mathbb{N}}$ extracted from $({\bf Q}_k)_{k\in\mathbb{N}}$ converges towards 
$\overline{{\bf Q}}_*$. For this, we denote by $\overline{{\bf Q}}_{\psi,*}$ the limit of the above subsequence, and prove that 
this matrix verifies property (\ref{eq:carac-max}). We first remark that sequences $\bs{\delta}^{\psi(k)+1}$ and $\tilde{\bs{\delta}}^{\psi(k)+1}$ converge towards vectors denoted $\bs{\delta}^{\psi,*}$ and
 $\tilde{\bs{\delta}}^{\psi,*}$ respectively. Moreover, $(\bs{\delta}^{\psi,*}, \tilde{\bs{\delta}}^{\psi,*})$
is solution of system (\ref{eq:canonique}) in which matrix ${\bf Q}$ coincides with $\overline{{\bf Q}}_{\psi,*}$. 
Therefore, using relations (\ref{eq:derivee-kappa}) and (\ref{eq:derivee-tildekappa}) as in the proof 
of Proposition \ref{prop:waterfilling}, we obtain that $\langle\nabla \overline{I}(\overline{{\bf Q}}_{\psi,*}), {\bf P} - \overline{{\bf Q}}_{\psi,*}\rangle$ coincides 
with $\langle\nabla_{{\bf Q}} {\cal V}(\overline{{\bf Q}}_{\psi,*}, \bs{\delta}_{\psi,*}, \tilde{{\bs \delta}}_{\psi,*}), {\bf P} - \overline{{\bf Q}}_{\psi,*}\rangle$. It remains to show that this term is negative for each ${\bf P}$ to complete the proof. For this, we use that 
${\bf Q}_{\psi(k)}$ is the argmax over ${\cal C}_1$ of function ${\bf Q} \rightarrow {\cal V}({\bf Q}, \bs{\delta}^{\psi(k)}, \tilde{\bs{\delta}}^{\psi(k)})$.
Therefore, 
\begin{equation}
\label{eq:derivee-negative}
\langle\nabla_{{\bf Q}} {\cal V}({\bf Q}_{\psi(k)}, \bs{\delta}_{\psi(k)}, \tilde{{\bs \delta}}_{\psi(k)}), {\bf P} - {\bf Q}_{\psi(k)}\rangle
\leq 0
\end{equation}
By (\ref{eq:hypothese}), sequences $(\bs{\delta}_{\psi(k)})_{k \geq 0}$ and $(\tilde{\bs{\delta}}_{\psi(k)})_{k \geq 0}$ converge towards 
$\bs{\delta}^{\psi,*}$ and  $\tilde{\bs{\delta}}^{\psi,*}$ respectively. Taking the limit of (\ref{eq:derivee-negative}) 
when $k\rightarrow\infty$ shows that $\langle\nabla_{{\bf Q}} {\cal V}(\overline{{\bf Q}}_{\psi,*}, \bs{\delta}_{\psi,*}, \tilde{{\bs \delta}}_{\psi,*}), {\bf P} - \overline{{\bf Q}}_{\psi,*}\rangle \leq 0$ as required. 
\end{IEEEproof}

To conclude, if the algorithm is convergent, that is, if the sequence of $({\bf Q}_k)_{k\in\mathbb{N}}$ converges towards a certain matrix, then the $\drl^{(k)}=\drl({\bf Q}_{k-1})$ and the $\dtl^{(k)}=\dtl({\bf Q}_{k-1})$ converge as well when $k\rightarrow\infty$.
(\ref{eq:hypothese}) is verified, hence,%
if the algorithm is convergent it converges towards $\overline{\bf Q}_*$.
Although the convergence of the algorithm has not been proved, this result is encouraging and suggests that the algorithm is reliable.
In particular, in all the conducted simulations the algorithm was converging.
In any case, condition (\ref{eq:hypothese}) can be easily checked.
If it is not satisfied, it is possible to modify the initial point ${\bf Q}_0$ as many times as needed to ensure the convergence.

\section{Numerical Results}
\label{sec:simulations}

\begin{figure}[!t]
 \centering
 \includegraphics[width=3in]{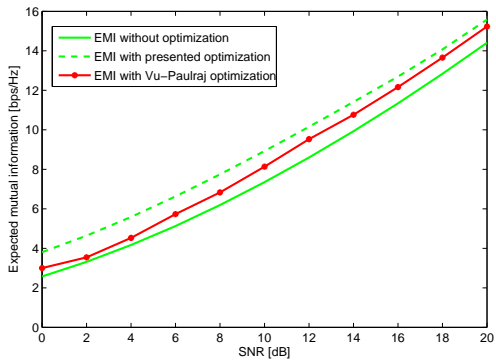}
 \caption{Comparison with Vu-Paulraj algorithm}
 \label{fig_comp}
\end{figure}

We provide here some simulations results to visualize the impact of the transmit correlation optimization in a realistic context. We use the propagation model introduced in \cite{boelcskei}, in which each path corresponds to a scatterer cluster characterized by a mean angle of departure and an angle spread.

In the featured simulations, we consider a frequency selective MIMO system with $t=r=4$, a carrier frequency of 2GHz, a number of paths $L=5$. The paths share the same power, and their mean departure angles and angles spreads are given in Table \ref{table_para} (in radians). In Figure \ref{fig_comp},
we have represented the true EMI $I({\bf I})$ (i.e. without optimization), and the optimized EMI $I(\overline{\bf Q}_*)$ (i.e. with an input covariance matrix maximizing the approximation). The EMI are evaluated by Monte-Carlo simulations, with $10^5$ channel realizations. The EMI optimized with Vu-Paulraj algorithm \cite{Vu-Paulraj-05} is also represented for comparison. We fixed to 10 the number of iterations in \cite{Vu-Paulraj-05}.

Figure \ref{fig_comp} shows that maximizing $\overline{I}({\bf Q})$ over the input covariance leads to significant improvement for $I({\bf Q})$.
Our approach provides the same results as Vu-Paulraj's algorithm at high SNR, and performs even better elsewhere. Vu-Paulraj's approach is penalized by the barrier method if the optimal input covariance is close to be singular, which is here the case if the SNR is not high enough. Moreover our algorithm is computationally much more efficient: in Vu-Paulraj's algorithm the evaluation 
of the gradient and of the Hessian of $I({\bf Q})$ needs heavy Monte-Carlo simulations ($10^4$ trials were used). Table \ref{table_comp} gives for both algorithms the average execution time in seconds to obtain the input covariance matrix, on a 3.16GHz Intel Xeon CPU with 8GB of RAM, for a number of paths $L=3$, $L=4$ and $L=5$.

\begin{table}[!t]
\renewcommand{\arraystretch}{1.3}
\caption{Paths angular parameters \normalfont{\scriptsize\it(in radians)}}
\label{table_para}
\centering
\begin{tabular}{|c|c|c|c|c|c|}
\cline{2-6}\multicolumn{1}{c|}{}  & $l=1$ & $l=2$ & $l=3$ & $l=4$ & $l=5$
\\\hline {mean departure angle} &$6.15$ & $3.52$ & $4.04$ & $2.58$ & $2.66$
\\\hline {departure angle spread} & $0.06$ & $0.09$ & $0.05$ & $0.05$ & $0.03$
\\\hline {mean arrival angle} &$4.85$ & $3.48$ & $1.71$ & $5.31$ & $0.06$
\\\hline {arrival angle spread} &$0.06$ &$0.08$ & $0.05$ & $0.02$ & $0.11$
\\\hline
\end{tabular}
\end{table}

\begin{table}[!t]
\renewcommand{\arraystretch}{1.3}
\caption{Average execution time \normalfont{\scriptsize\it(in seconds)}}
\label{table_comp}
\centering
\begin{tabular}{|c|c|c|c|}
\cline{2-4}\multicolumn{1}{c|}{} & $L=3$& $L=4$& $L=5$\\
\hline
Vu-Paulraj & $903$& $1245$& $1649$\\
\hline
New algorithm & $7,0.10^{-3}$& $7,4.10^{-3}$& $8,3.10^{-3}$\\
\hline
\end{tabular}
\end{table}

\section{Conclusion}
In this paper we have addressed the evaluation of the capacity achieving covariance matrices of frequency selective MIMO channels. We have proposed to optimize a large system approximation of the EMI, and have introduced an attractive iterative algorithm.
\bibliographystyle{IEEEtran}
\bibliography{IEEEabrv,bibMarne}

\begin{thebibliography}{10}
\providecommand{\url}[1]{#1}
\csname url@samestyle\endcsname
\providecommand{\newblock}{\relax}
\providecommand{\bibinfo}[2]{#2}
\providecommand{\BIBentrySTDinterwordspacing}{\spaceskip=0pt\relax}
\providecommand{\BIBentryALTinterwordstretchfactor}{4}
\providecommand{\BIBentryALTinterwordspacing}{\spaceskip=\fontdimen2\font plus
\BIBentryALTinterwordstretchfactor\fontdimen3\font minus
  \fontdimen4\font\relax}
\providecommand{\BIBforeignlanguage}[2]{{%
\expandafter\ifx\csname l@#1\endcsname\relax
\typeout{** WARNING: IEEEtran.bst: No hyphenation pattern has been}%
\typeout{** loaded for the language `#1'. Using the pattern for}%
\typeout{** the default language instead.}%
\else
\language=\csname l@#1\endcsname
\fi
#2}}
\providecommand{\BIBdecl}{\relax}
\BIBdecl

\bibitem{Goldsmith-Jafar-etal-03}
A.~Goldsmith, S.~Jafar, N.~Jindal, and S.~Vishwanath, ``{Capacity limits of
  MIMO channels},'' \emph{{IEEE} J. Select. Areas Commun.}, vol.~21, no.~5, pp.
  684--702, 2003.

\bibitem{Hoesli-Kim-Lapidoth-05}
D.~Hoesli, Y.~Kim, and A.~Lapidoth, ``{Monotonicity results for coherent MIMO
  Rician channels},'' \emph{{IEEE} Trans. Inform. Theory}, vol.~51, no.~12, pp.
  4334--4339, 2005.

\bibitem{boelcskei}
H.~Bolckei, D.~Gesbert, and A.~Paulraj, ``{On the capacity of OFDM-based
  spatial multiplexing systems},'' \emph{{IEEE} Trans. Commun.}, vol.~50,
  no.~2, pp. 225--234, 2002.

\bibitem{Moustakas-Simon-07}
A.~Moustakas and S.~Simon, ``{On the outage capacity of correlated
  multiple-path MIMO channels},'' \emph{{IEEE} Trans. Inform. Theory}, vol.~53,
  no.~11, p. 3887, 2007.

\bibitem{Vu-Paulraj-05}
M.~Vu and A.~Paulraj, ``{Capacity optimization for Rician correlated MIMO
  wireless channels},'' in \emph{Proc. Asilomar Conference}, 2005, pp.
  133--138.

\bibitem{Wen-Com-06}
C.~Wen, P.~Ting, and J.~Chen, ``{Asymptotic analysis of MIMO wireless systems
  with spatial correlation at the receiver},'' \emph{{IEEE} Trans. Commun.},
  vol.~54, no.~2, pp. 349--363, 2006.

\bibitem{dum-hac-las-lou-naj-09}
J.~Dumont, W.~Hachem, S.~Lasaulce, P.~Loubaton, and J.~Najim, ``{On the
  capacity achieving covariance matrix for Rician MIMO channels: an asymptotic
  approach},'' \emph{{IEEE} Trans. Inform. Theory}, vol.~56, no.~3, pp.
  1048--1069, 2010.

\bibitem{Cedric-GC}
C.~Artigue, P.~Loubaton, and B.~Mouhouche, ``On the ergodic capacity of
  frequency selective {MIMO} systems equipped with {MMSE} receivers: An
  asymptotic approach,'' in \emph{Proc. Globecom}, 2008.

\bibitem{hac-kor-lou-naj-pas-08}
W.~Hachem, O.~Khorunzhiy, P.~Loubaton, J.~Najim, and L.~Pastur, ``{A new
  approach for capacity analysis of large dimensional multi-antenna
  channels},'' \emph{{IEEE} Trans. Inform. Theory}, vol.~54, no.~9, 2008.

\bibitem{HLN07}
W.~Hachem, P.~Loubaton, and J.~Najim, ``{Deterministic equivalents for certain
  functionals of large random matrices},'' \emph{Annals of Applied
  Probability}, vol.~17, no.~3, pp. 875--930, 2007.

\bibitem{horn1990matrix}
R.~Horn and C.~Johnson, \emph{{Matrix analysis}}.\hskip 1em plus 0.5em minus
  0.4em\relax Cambridge Univ Pr, 1990.

\bibitem{horn1994topics}
------, \emph{{Topics in matrix analysis}}.\hskip 1em plus 0.5em minus
  0.4em\relax Cambridge Univ Pr, 1994.

\bibitem{dumont-2007}
\BIBentryALTinterwordspacing
J.~Dumont, W.~Hachem, S.~Lasaulce, P.~Loubaton, and J.~Najim, ``On the capacity
  achieving covariance matrix for {Rician MIMO} channels: an asymptotic
  approach,'' 2007. [Online]. Available: \url{http://arxiv.org/abs/0710.4051}
\BIBentrySTDinterwordspacing

\end{thebibliography}
\end{document}